\documentclass[twocolumn,showpacs]{revtex4}
\usepackage{graphics,epsfig,graphicx}
\usepackage{color}
\usepackage{makeidx}
\usepackage{latexsym}

\begin{document}


\title{Bandwidth-tunable single photon source in an ion trap quantum network}

\author{M. Almendros, J. Huwer, N. Piro, F. Rohde, C. Schuck, M. Hennrich, F. Dubin, J. Eschner}
\affiliation{ICFO-Institut de Ci\`{e}ncies Fot\`{o}niques, Mediterranean Technology Park, E-08860 Castelldefels (Barcelona), Spain}
\date{\today}

\begin{abstract}
We report a tunable single-photon source based on a single trapped ion. Employing
spontaneous Raman scattering and in-vacuum optics with large numerical aperture, single
photons are efficiently created with controlled temporal shape and coherence time. These can be varied between 70~ns and 1.6~$\mu$s, as characterized by operating two sources simultaneously in two remote ion traps which reveals mutual and individual coherence through two-photon interference.
\end{abstract}

\maketitle

In current designs of quantum networks based on individual quantum systems, controlled
emission and absorption of single photons by single atomic systems serves for
implementing local quantum memories. It also allows one to transport quantum information and
establish entanglement between distant locations, or nodes \cite{Moehring_07}. The latter
operation is based on the underlying entanglement between a quantum system and the
photons it emits \cite{Blinov_04}. It also employs single- or two-photon interference as an essential resource \cite{Zippilli_08}: the coincident detection of
two photons emitted by distinct quantum nodes and interfering on a beam splitter can
project the nodes into an entangled state \cite{Simon_03}.
Alternatively, entanglement may be generated through two indistinguishable scattering
paths leading to a single photon detection \cite{Cabrillo_99}. In both approaches, the
coherence of individual photons is a critical ingredient.  In fact, it controls the visibility of any two-photon interference through the temporal overlap of incident photon wavepackets \cite{HOM}. For interfering broadband photons, high fidelity entanglement of quantum nodes thereby require accurate synchronization of photon emissions.  On the other hand, for gate operations based on single-photon interference, the photon coherence sets the maximum path length difference between quantum nodes at the detectors.

Besides in quantum networks, single-photon sources are important
building blocks in schemes of quantum communication \cite{Cirac_97},
cryptography \cite{Beveratos_02}, and optical quantum computing
\cite{Knill_01}. The coherence of the photons is particularly
important for optimising the efficiency of quantum information
transmission, the ideal situation being tunable and Fourier transform-limited
single photons which make optimal use of the available transmission
bandwidth.

Single-photon sources have so far been implemented with various
systems which act as functional quantum memories, for instance with
quantum dots and defects, neutral atoms and trapped ions (see
\cite{Grangier_04} for a review). With single-atom devices, two
approaches may be distinguished: on the one hand, single atoms (or
ions) strongly coupled to high finesse resonators allow one to
efficiently release individual photons with long coherence into a
single optical mode \cite{McKeever_04, Keller_04,
Hiljema_07,Barros_09}. Quantum gates based on single- or two-photon interference can hence be implemented 
with flexible constraints, however, at the cost of a controlled interaction between the atom/ion and the cavity optical mode. On the other hand, single atoms or ions
trapped in free space may be employed. Compared to atom-cavity
devices, these in general have a multi-mode emission pattern which reduces the fraction of photons that can be collected in a single optical mode. Resonance fluorescence photons are furthermore broadband, with coherence times generally less than 10 ns, but exhibit a high degree of indistinguishability \cite{Dubin_07,Beugnon_06, Maunz_07, Gerber_09}. Hence, two remotely trapped ions have been entangled based on the coincident detection of two photons \cite{Moehring_07b}. Moreover, note that trapped atomic ensembles also allow one to herald single
photons at high efficiency \cite{Chou_04} with a corresponding
bandwidth which possibly reaches the Fourier limit
\cite{Thompson_06}.

In this letter, we present experiments where single ions are trapped in front of high numerical aperture laser objectives. Such experimental setup constitutes a tunable and narrowband single-photon source that operates with an efficiency comparable to what is reported with ion-cavity devices \cite{Keller_04,Barros_09}. Emitted photons have a coherence which is revealed by measuring time-resolved quantum 
interference between photons emitted by two ions in
distant traps. Varying the photon degree of indistinguishability, we observe Hong-Ou-Mandel type of interference \cite{HOM} and deduce a coherence time approaching the Fourier limit. Our analysis shows that single ions trapped in front of high numerical laser objectives are relevant candidates to implement the nodes of a network with multiple and narrowband quantum communication channels.

\begin{figure}[ht]
\includegraphics[width=\columnwidth]{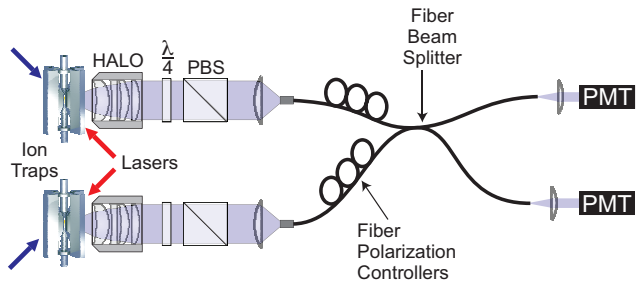}
\caption{(Color Online) Schematic representation of the experimental setup. It comprises
two linear Paul traps confining single ions. The ions are cooled and excited by the same
lasers (solid arrows) at 397~nm and 866~nm, and their blue fluorescence is collected by
two high numerical aperture objectives (HALOs) per trap (one objective is used for camera
imaging and is not shown). The polarization of detected photons is selected by a
combination of a quarter wave-plate ($\lambda/4$) and a polarizing beam splitter (PBS).
The light from the two ions is superimposed on a fiber beam splitter after further
polarization control, represented by the loops. At the beam splitter output, single
photons are detected by photomultiplier tubes (PMT) with $\sim 1$~ns time resolution. A
constant magnetic field along the trap-HALO direction serves as quantization axis while
stray magnetic fields are compensated.}
\end{figure}

The experimental setup is presented in Figure 1. Single $^{40}$Ca$^+$ ions are trapped in
two vacuum vessels separated by approximately one meter \cite{Gerber_09}. They are
excited and cooled on their S$_{1/2}-$P$_{1/2}$ and D$_{3/2}-$P$_{1/2}$ electronic
transitions (see Fig.\ 2) by frequency-stabilized lasers at 397~nm (blue) and 866~nm
(infrared), polarized perpendicularly to the quantization axis. About $4\%$ of the blue
fluorescence photons are collected by in-vacuum high numerical aperture objectives
(HALOs) and coupled into the two input arms of a single-mode fiber beam splitter.
Single-photon detectors at the beam splitter outputs record individual photon arrival
times, from which correlations are computed.

Figure 2 presents an example of laser excitation sequence which we use to create single
photons. It actually corresponds to the one utilized for the experiments shown in Fig. 4, and consists of alternating periods of laser cooling and manipulation of the ion's internal state. Excitation parameters such as laser intensities and detunings, as well as the
durations of the periods of the sequence, are optimized by performing numerical
simulations. The latter are based on time-dependent optical Bloch equations describing
the dynamics of the 8 relevant electronic levels of the single-ion wave function
\cite{Toschek}.

\begin{figure}[h]
\includegraphics[width=.9\columnwidth]{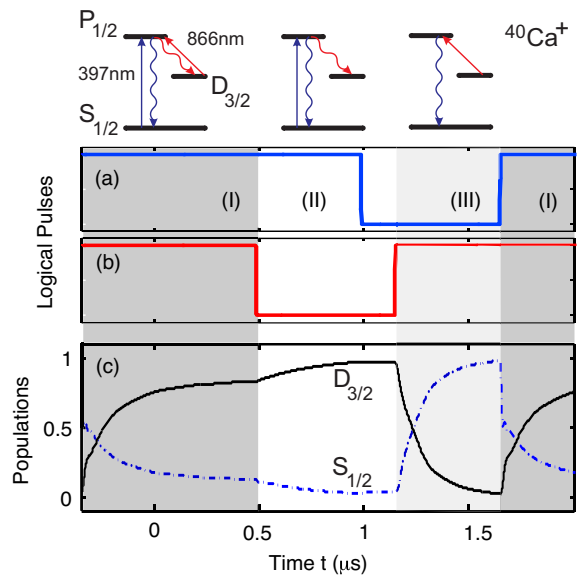}
\caption{(Color Online) Schematic representation of the switching of (a) pump (397~nm)
and (b) repump (866~nm) laser during an excitation sequence. In the top panel a
simplified level scheme of $^{40}$Ca$^+$ is presented together with the excitation lasers
and spontaneous decay channels. Note that this sequence is utilized in the experiments reported in Fig. 4. (c) Calculated time evolution of the populations in the
S$_{1/2}$ and D$_{3/2}$ electronic manifolds for typical experimental parameters.}
\end{figure}

In Figure 2c, an example for the evolution of populations in the S$_{1/2}$ and D$_{3/2}$
manifolds during one excitation sequence is presented. The cooling period (part I), with
both infrared "repump" and blue "pump" laser on, has a minimum duration (1 $\mu$s) such that a steady state
is closely approached at its end. This ensures that the motion of the ion is efficiently
cooled, before projecting the internal state into the D$_{3/2}$ manifold by switching off
the repump laser (part II). This state preparation is achieved with as high as 95~$\%$
efficiency. Thereafter, the repump laser is turned on again while the pump laser has
previously been switched off (part III). This transfers the electronic population back to
the S$_{1/2}$ state and releases a blue photon.

We control the rate $\Gamma$ of this spontaneous Raman transition via the intensity of
the repump laser ($\propto \Omega^2_\mathrm{r}$, where $\Omega_\mathrm{r}$ is the
infrared Rabi frequency) and its detuning $\Delta_\mathrm{r}$. In the regime of our experiment, $\Delta_\mathrm{r} \sim \gamma_{\mathrm P}$ and
$\Omega_\mathrm{r} \lesssim \gamma_{\mathrm P}$, $\gamma_{\mathrm P}=2\pi\times24$~MHz being
the decay rate of the P$_{1/2}$ level, one expects a linear variation of $\Gamma$ as a
function of the repump intensity.

\begin{figure}[h]
\includegraphics[width=.9\columnwidth]{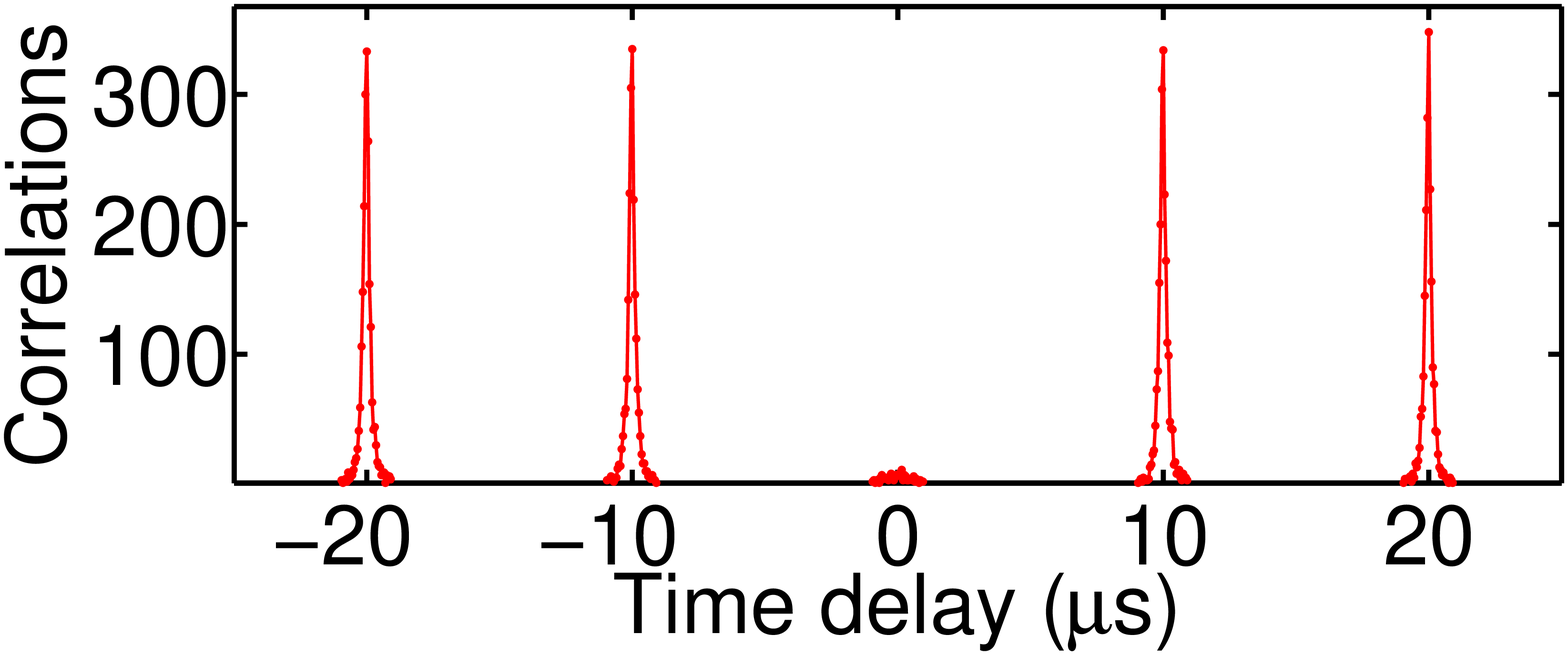}\vspace{.1cm}
\includegraphics[width=0.9\columnwidth]{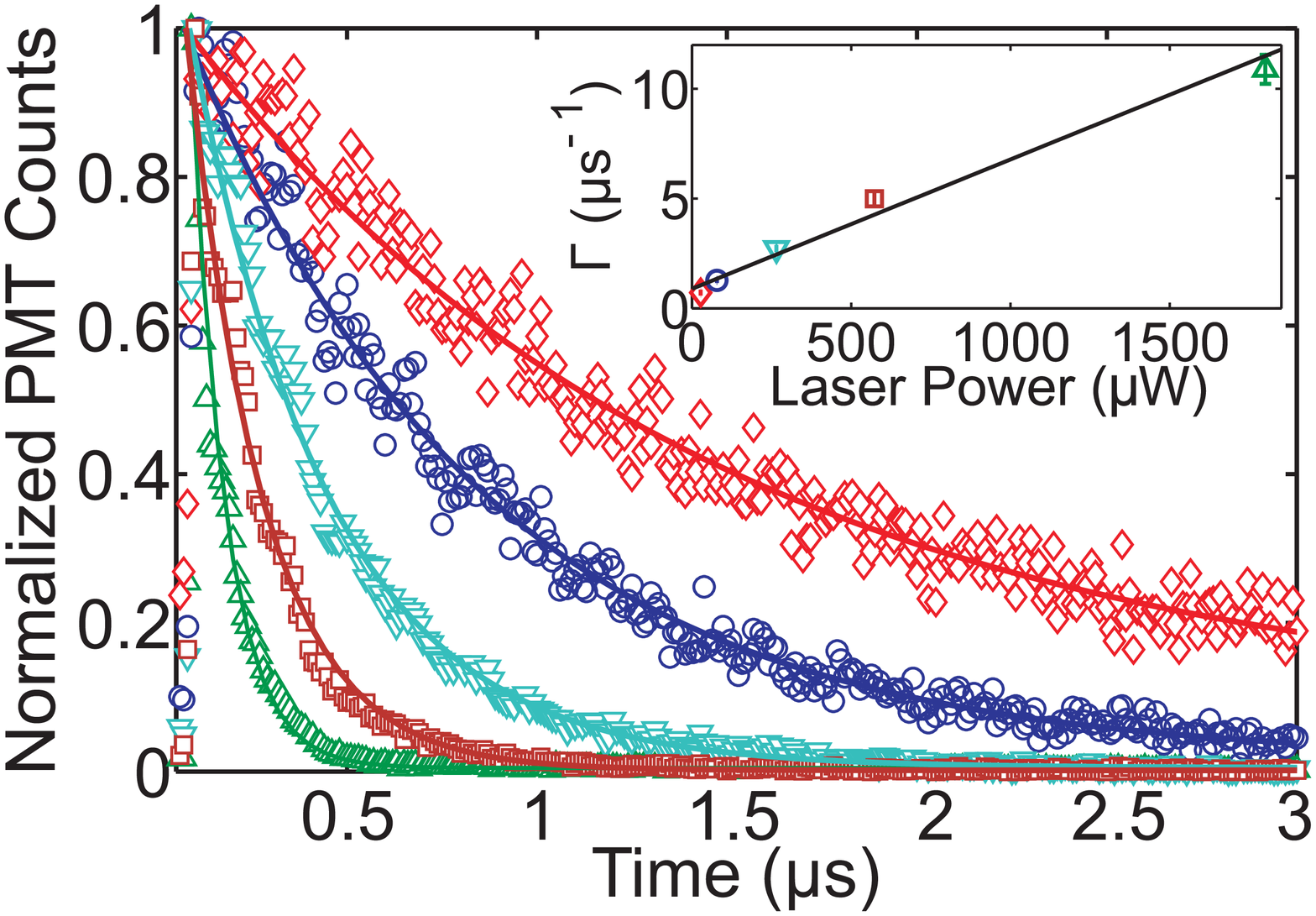}
\caption{(Color Online) Characterization of one individual single-photon source. (a)
Intensity auto-correlation of the emitted light. Measurements are displayed with 50 ns time resolution and correspond to 15 min acquisition. (b) Distributions of single-photon
detection times for various values of the repump laser intensity. Experimental data are displayed with 10 ns time resolution while each curve correspond to 5 min acquisition.  Inset: Raman transition
rate $\Gamma$ as a function of the repump laser intensity.}
\end{figure}

The first set of experimental results, displayed in Fig.~3, characterizes one individual
single-photon source implemented in one trap. Here the sequence period is set to
$10~\mu$s, and photons are collected into multi-mode optical fibers for higher signal to background ratio ($\approx 200$). The measured detection probability per
excitation pulse is $4 \times 10^{-3}$. The photodetectors are gated, such that only
photons emitted during the repump phase (part (III) in Fig.~2) are counted. Figure 3a shows
the resulting second-order correlation of the arrival times of photons emitted by a
single ion. Most importantly, one notes a nearly vanishing peak around zero delay time.
This demonstrates that at most a single photon is emitted while exciting the spontaneous
Raman transition from D$_{3/2}$ to S$_{1/2}$. The observed residual signal agrees with
the expected rate of accidental correlations between emitted photons and dark counts of
the photodetectors. For longer time delays, the number of correlations vanishes when the
detectors are gated off, while subsequent peaks signal correlations between photons
emitted during successive sequences.

In Fig.~3b we present the distribution of single photon detection times, corresponding to
the shape of the single-photon wave packet \cite{Keller_04}. It is obtained by
correlating photon arrival times with the logical trigger gating on the photodetectors,
i.e.\ with the beginning of the repump phase. In these measurements we varied the
intensity of the repump laser, $\propto \Omega_r^2$, for a fixed detuning
$\Delta_\mathrm{r}/2\pi\approx -55(5)$~MHz from the D$_{3/2}-$P$_{1/2}$
resonance frequency. The wave packet is characterized by a steep initial rise and a
subsequent exponential decay, whose time constant $T_1=1/\Gamma$ we continuously control between 70~ns
and 1.6~$\mu$s. The inset shows that the respective rate $\Gamma$ varies linearly with
the excitation power, as expected. For stronger excitation one encounters more complex
dynamics where the infrared laser induces coherent coupling between the D$_{3/2}$ and
P$_{1/2}$ manifolds, signaled by damped Rabi oscillations in the arrival times of blue
photons (see Fig. 4a).

The experiments reported in Figure 3 show how we operate a single ion as a clean source
of single photons emitted at a tunable rate, from nanosecond to microsecond timescales.
Furthermore, light collection by the HALOs allows us to efficiently couple the generated
photons into a single optical mode without the use of a cavity: with a wave packet width
of $T_1\approx 250$~ns and a sequence repetition rate of 500~kHz, we reach a single mode
fiber-coupled detection rate of $\approx 200$~photons/s using a single HALO ($\approx
1000$/s before the detector).

In order to assess also the coherence time of our single-photon source, we implemented a
primitive quantum network scenario, operating two sources with two single ions in
independent traps (see Fig.~1). We then studied the quantum interference of two photons
emitted by the distant ions. Two-photon interference, as originally shown by Hong, Ou,
and Mandel, consists in the coalescence of two identical photons impinging at the two
input ports of a beam splitter \cite{HOM}. Both photons emerge in the same output port,
inducing a dip in the coincidence rate, whose shape characterizes the coherence of the
interfering light fields \cite{Legero_04, Legero_longer}.

A fundamental prerequisite for two-photon interference is indistinguishability of the
photons. This controls the interference contrast, {\it i.e.} the amplitude of the dip in
the two-photon correlations. Hence, one must aim at ideal spatial mode matching as well
as identical polarizations at the beam splitter. In our experiments this is guaranteed by
the combination of a fiber beam splitter together with fiber polarization controllers.
Furthermore, the frequency of the interfering photons has to be controlled accurately.
This is achieved by deriving the exciting light from the same laser and by adjusting the
magnetic fields in the two traps to the same value. Emitted blue photons are collected along the quantization axis. A combination of
quarter-wave plate and polarizing beam-splitter ensures that solely right-circularly
polarized photons are coupled into the fibers. Then, with the repump laser polarized
perpendicularly to the magnetic field, electronic populations prepared in the $m=-3/2$
and $m=+1/2$ magnetic sub-states of the D$_{3/2}$ manifold contribute to the detected
spontaneous Raman emissions. Due to the Zeeman splitting induced by a magnetic field of
$B=2.2(1)$~G, incident photons can exhibit a frequency difference of $2g_{\mathrm D}
\mu_{\mathrm B} B= 7.5(4)$~MHz.

Figure 4 presents results from the two networked single-photon sources. In Fig.~4a the
normalized time distributions of emitted photons, $n(\tau)$, are shown for both
sources. These were adjusted experimentally such that both ions yield near-identical
profiles. One notes damped Rabi oscillations on the D$_{3/2}$ to P$_{1/2}$ transition,
induced by the repump laser intensity being close to saturation. Figure~4b presents the
normalized second-order correlation between the two detectors when photons emitted by the
two ions are distinguishable, $g_{ni}^{(2)}(\tau)$. Experimentally, this is achieved by
making the polarizations at the beam splitter orthogonal. From the width of the central
peak in $g_{ni}^{(2)}(\tau)$ we recover the width of the single-photon wave-packet, $T_1
=$ 230(30) ns.

\begin{figure}[h]
\includegraphics[width=0.925\columnwidth]{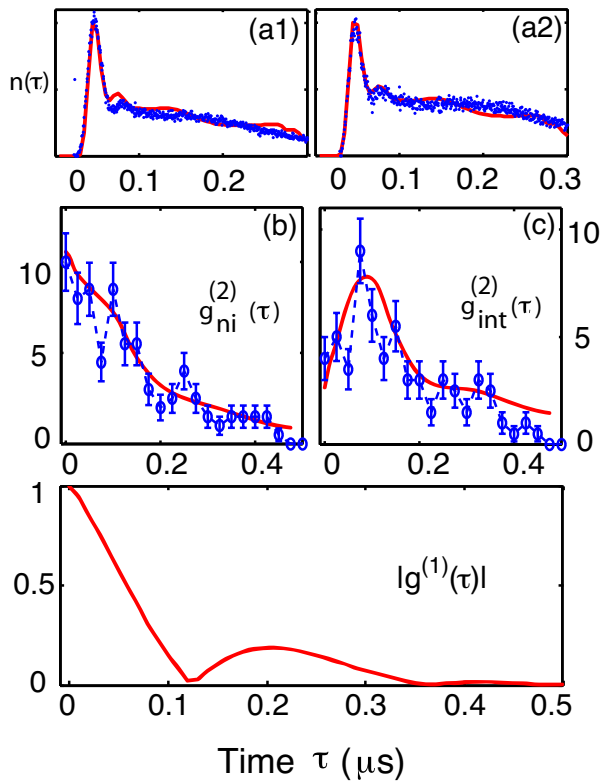}
\caption{(Color Online): Results with two identical single-photon sources in distant
traps. (a1) and (a2): Detection time distributions of photons emitted from each trap. (b)
and (c): Normalized two-photon correlations in 25~ns time bins for distinguishable (b)
and indistinguishable (c) photons. (d): Coherence of individual light fields deduced from
(b) and (c). Solid lines represent the result of our theoretical model obtained with
calibrated experimental parameters, while experimental data are displayed with Poissonian statistics at all times.}
\end{figure}

In Fig.~4c, emitted photons are rendered indistinguishable by adjusting their
polarizations to be along the same axis. The normalized correlation function
$g_{int}^{(2)}(\tau)$ exhibits a dip for short time delays as expected for two-photon interference. We observe a 50(20)$\%$ suppression of coincidences, i.e interference contrast, without subtracting accidental correlations. 

Solid lines in Fig.~4 show the results of our theoretical calculations. We utilize the
model described in more detail in \cite{Gerber_09}, based on 8-level atomic dynamics and
independently determined experimental parameters. We also include experimental limitations such as the finite optical switching of acousto-optics modulators used to create laser pulses. Indeed, for the repetition rate used in the experiments of Fig. 4 (500 kHz), we observe that a small fraction (5$\%$) of cooling (blue) laser-light persists for 200 ns during part (III) of the sequence shown in Figure 2. This alters the purity of our single-photon sources as it induces two-photon emission by individual ions.

In Figure 4, quantitative agreement is obtained for the time distributions of photon detection from both trap (see panels (a1) and (a2)). Furthermore, second-order correlations between photons with orthogonal (Fig.~4b) and equal polarizations (Fig.~4c) coincide with our theoretical expectations within the statistical precision. Noting that correlations between detector dark counts and signal photons solely induce 5 and 8 $\%$ of correlations in the experiments (b) and (c) respectively, we conclude that the number of coincidences for indistinguishable photons (panel (c)) arises from correlations between photons emitted by the same ion.  In future works, these events will be suppressed by improved optical switching leading to very high interference contrast.

As a final result, we present in Fig.~4d the first-order coherence of emitted photons, $g^{(1)}(\tau)$, deduced
from our theoretical model. It exhibits an oscillatory behavior which we interpret as arising from
the beat of the two photon frequencies of our source and the strong coherent excitation of the D$_{3/2}-$P$_{1/2}$ transition by the repump infrared laser. Furthermore, one notes an approximately exponential envelope with decay time $\sim 250$~ns, which reveals the coherence time $T_2$ of
the individual photons. An optimally coherent free-space single-photon source would
feature a coherence time $T_2=2T_1$; apart from the modulation due to the two Raman
transition channels, with $T_2\geq T_1$ we are approaching this limit.

To conclude, we control and manipulate the multi-level internal structure of single ions
to flexibly and efficiently engineer two nodes of a primitive quantum network, emitting
photons with controlled temporal and coherence properties. In future developments,
single-frequency operation will be implemented, for which individual state addressing \cite{Nagerl_2000} or dark-resonance \cite{Champenois_06}  will be employed during the pump phase. For
enhanced collection efficiency, both laser objectives will be used, including enhancement
by constructive single-photon interference \cite{Eschner2001}, and the single-mode fiber
coupling will be improved. At least 10-fold efficiency improvement is expected hence yielding overall efficiencies remarkably higher than reported with cavity-based single photon sources \cite{Keller_04,Barros_09}.

We acknowledge support from the European Commission (SCALA, contract 015714; EMALI,
MRTN-CT-2006-035369), the Spanish MICINN (QOIT, CSD2006-00019; QLIQS, FIS2005-08257;
QNLP, FIS2007-66944), and the Generalitat de Catalunya (2005SGR00189; FI-AGAUR fellowship
of C.S.).

\end{document}